E.Loza, V.Vaschenko

# Madelung rule violation statistics and superheavy elements electron shell prediction

*State Ecological Academy, Kyiv, Ukraine. E-mail:* *Loza@bmyr.kiev.ua*

**Abstract**

*The paper presents tetrahedron periodic table to conveniently include superheavy elements. Madelung rule violation statistics is discussed and a model for Madelung rule violation probability calculation is proposed. On its basis superheavy elements probable electron shell structure is determined.*

## Introduction

The electron shell structure was experimentally investigated for 104 elements. More heavy elements with atomic numbers 105 to 118 have been synthesized [1] but not yet investigated. Some approximate quantum mechanics calculations have been made for these elements [2,3] but it is a very complicated many-body quantum problem.

The true optical and chemical properties of the element may be determined only by experimental means, thou we may still make some suggestions on electron shell structure of superheavy nuclei based on Madelung rule. However, Madelung rule is not a strict rule and 21 from 104 known elements (20.2%) violate Madelung rule.

In the present paper we consider statistics of Madelung rule violation of known elements and its extrapolation to superheavy elements. Therefore such extrapolation is not based on theoretical quantum mechanics problem solution, but on analysis of 'experimental' data and its extrapolation.

Data on electron shell of known elements was obtained from http://en.wikipedia.org/wiki/Electron_configurations_of_the_elements_(data_page)

## 1. Tetrahedron periodic table and Madelung rule violation

There are many new variants of periodic table extension to include superheavy and yet anticipated but unsyntheised chemical elements. In the present paper we propose Tetrahedron variant of elements arrangement. The Idea of its creation is the following:

1. Elements of same electron shell must be placed together (e.g. d elements row must not be broken).

2. Elements with a same number of electrons must be placed together (up to "half filled electron shell" (middle) element).

3. Elements with a same number of electron deficiency must be placed together (down to the middle element).

4. The elements with half filled electron shell must be placed in the middle.

5. The elements with completely filled electron shell must be placed separately.

6. Inert gases must be places separately.

7. The elements with a specific number of electrons must be placed opposite of elements with the same number of electron deficiency.

8. Helium must occupy the cell for both of s2 and p6 elements.

It is convenient to present it in the following form (fig.1).

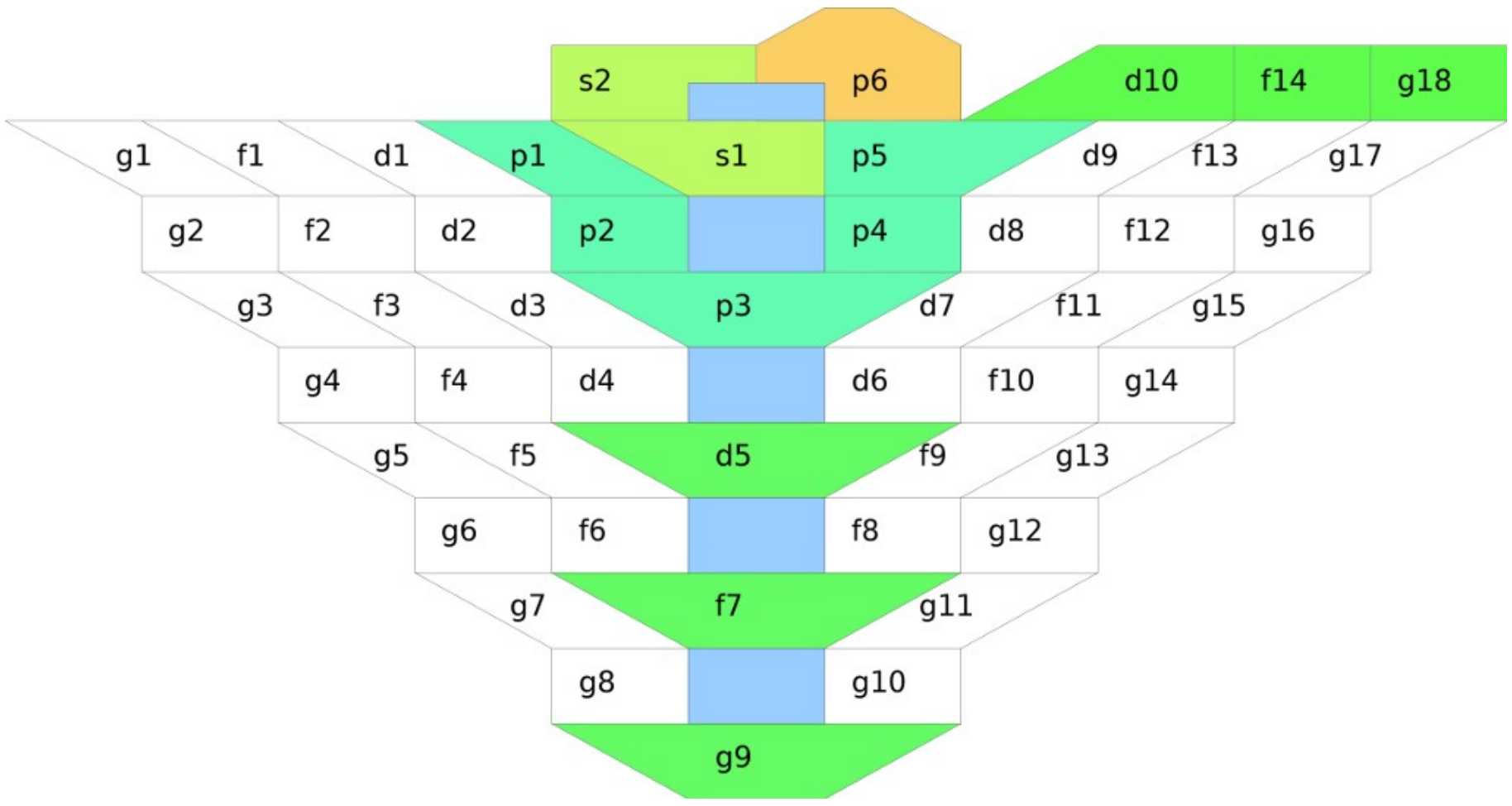

Fig.1. The principle of tetrahedron periodic table.

One more note, we consider the element to be located in the period, where the shell being filled is located. I.e. the periods of tetrahedron periodic table are not equal to periods of Mendeleev periodic table. And there will be no homogenity in atomic number increase in a single period.

We consider the total number of elements to be 137-138 according to relativistic Dirac equation calculation for ground state of superheavy elements [4]. In this case we obtain total of 8 periods. Note, that we name all elements with uninvestigated electron shell not by their proper name (which is often its atomic number in Latin), but by their atomic number.

The complete drawing of the tetrahedron periodic table periods is seen in fig.2. It is obvious that the tetrahedron periodic table is 3D.

Here let's note that it is almost symmetrical if we consider the total number of elements to be 138. It would have been completely symmetrical if the maximum number of elements was 220.

The 1st period is considered unique because helium is both s2 element and and an inert gas. Considering symmetry we suggest that 120th element may also have properties of helium. However, this is no more than a symmetry suggestion. Even in case the maximal number of elements is really 137, the p-shell shall also be empty and therefore 120th element should not be inert.

At fig.2 we distinguish three types of Madelung rule violation:

1. Simple Madelung rule violation
2. Double Madelung rule violation
3. Complex Madelung rule violation

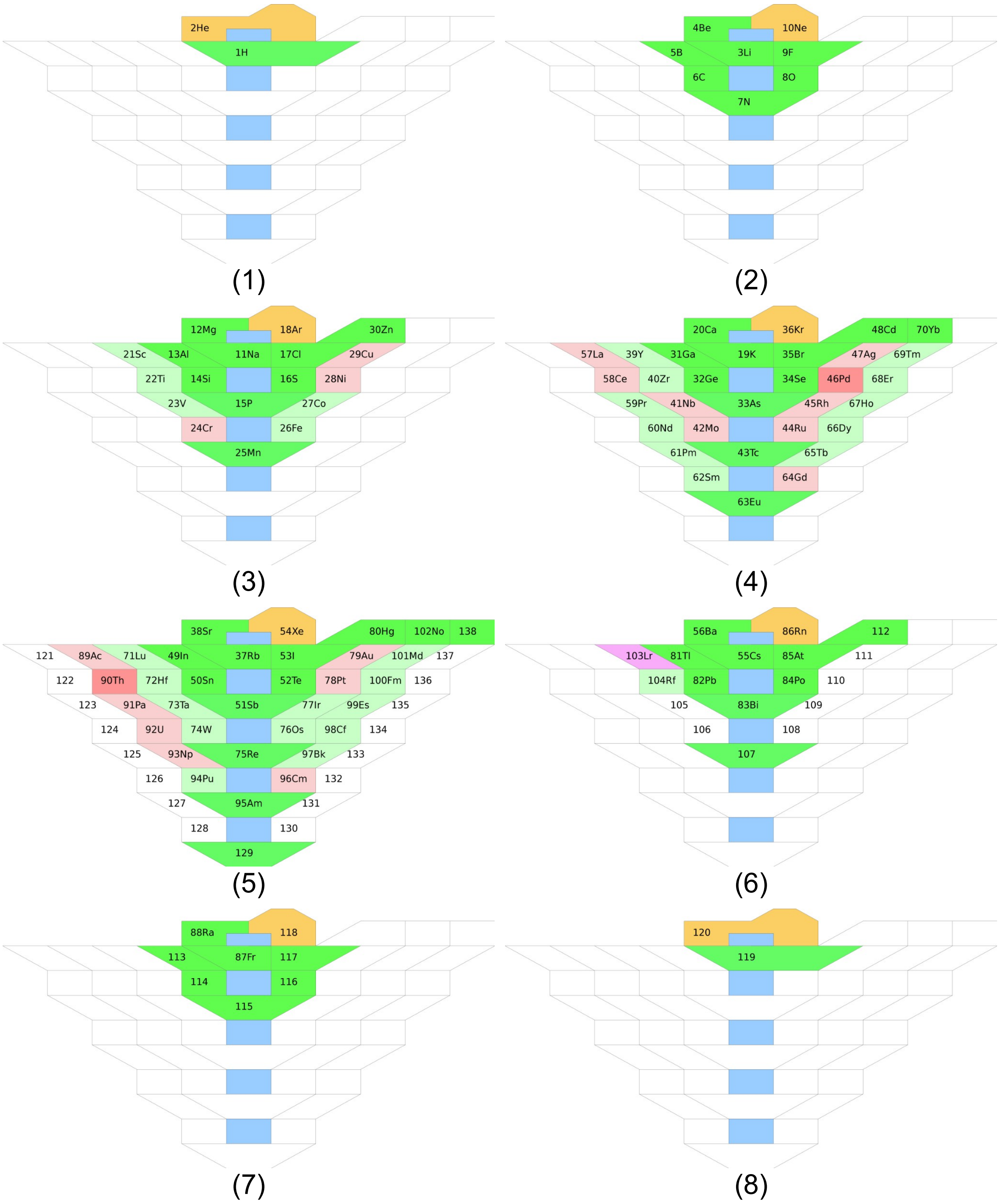

Fig.2. The tetrahedron periodic table up to the 138th element: (1)-(8) show the corresponding 1 to 8 periods of the periodic table; elements violating Madelung rule are presented in soft red color; double Madelung violation is given in intense red color and complex Madelung rule violation is given in purple; elements following Madelung rule are given in soft green color and unknown elements and unfilled table cells are white; inert gases are given in brown color; the intense green color signifies the extended rules obtained in the next chapter.

**Simple Madelung rule violation** means that a single electron moves to d-eletron shell (i.e. 1 s-electron descends to a corresponding one level lower d-shell for d-elements, and 1 f-electron ascends to a higher d-shell for f-elements). Simple Madelung rule violation is dominant and found in 10 of 26 (38.5%) of known d-elements and 8 of 24 (33.3%) of known f-elements.

Let us note that in the present paper we consider that in case of simple Madelung rule violation g-electrons will move to upper f-electron shell analogously to f-elements. Thou, this may not be quiet correct.

**Double Madelung rule violation** obviously means two electrons violating Madelung rule, i.e. descending to the lower electron shell. This is very rare and is found only in 1 of 26 (3.8%) d-elements and 1 of 24 (4.1%) f-elements. Let us note that Double Madelung rule violation is always preceded (and, probably, followed) by a simple violation (i.e. double violation for 90Th is preceded by simple violation in 58Co and double violation for 46Pd (these are f2 elements) is preceded by simple violation in 28Ni and followed by simple violation in 78Pt (these are d5 elements)).

Finally, **complex Madelung rule violation** is found only for unique d1-element and it is 103Lr (that is 3.8% for d-elements and 2% for all known d and f elements). It has unique electron structure $5f^{14}\ 7s^{2}\ 7p^{1}$ instead of $5f^{14}\ 6d^{1}\ 7s^{2}$. However this may be manifestation of a more complex violations with atomic number increase.

Together such non-simple violations total 3 elements out of 50 d and f elements, i.e. 6%.

It would have been very interesting to investigate electron shell structure depending on isotope number of the nuclei, however, such survey is outside of the present paper.

## 2. Extended rules for Madelung rule violation

The statistics of Madelung rule violation for known elements (later referred as 'experimental' data) is presented in fig.3. Note some kind of symmetry between s-elements and p-elements, and antisymmetry between d and f-elements.

As we can see in fig.3 some elements always obey to Madelung rule, and some elements always violate Madelung rule. We may suggest, that this image is not just statistics, but a part of a more complex electron shell structure rule.

I.e. we can see that p and s elements never violate Madelung rule. This statement is true for 10 s elements and 30 p elements (including inert gases). Therefore we may expect that further s and p elements will also obey Madelung rule exactly and formulate the following rule:

**SP-rule: s and p elements never violate Madelung rule**

As we can see from fig.3 all 10 elements (or 6 in case not including inert gases; or if both inert gases and 2s elements are included this rule is obeyed by 17 elements) with complete electron shell obeys Madelung rule. Therefore we may suggest the following rule:

**C-rule: every closed-shell element obeys Madelung rule.**

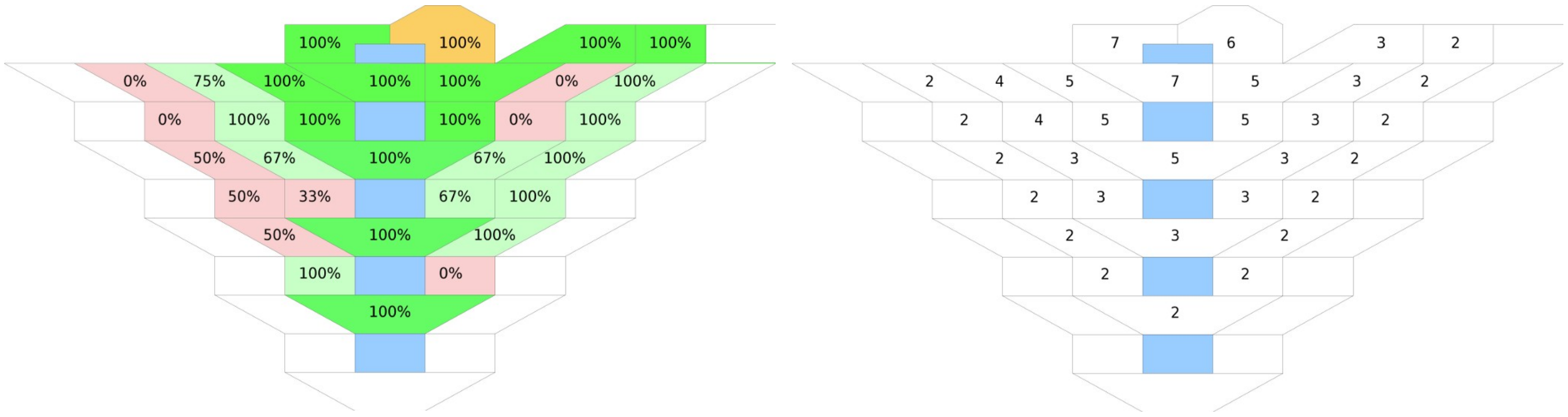

(a) (b)

Fig.3. Probability of Madelung rule obeying for known elements (a) and number of known elements per cell (b) (i.e. the number of events determining statistics reliability); intense green color - the extended rules (except 8-rule discussed later); soft red color - values of 0..50%; soft green color - values 51%..100%.

The next rule will deal with the 'middle' element. I.e. 3rd for p-elements (we may also include them into this rule), 5th for d-elements and 7th for f-elements. As we can see from fig.3 they also never violate Madelung rule. There are 10 such known elements. If we include 1s elements, their number will increase to 17.

Therefore the third rule for the middle element is:

**M-rule: Middle elements never violate Madelung rule.**

There are also some elements that always violate or always obey Madelung rule, thou we cannot identify their position by simple rules given above. Moreover they are too few to form reliable statistics.

For d-shell these non-strict rules are:

* **All 4 known d2 elements obey Madelung rule** (explained later by 8-rule given below): 22Ti, 40Zr, 72Hf, 104Rf.

* **All 3 known d8 elements violate Madelung rule** (explained later by 8-rule given below): 28Ni, 46Pd (double violation), 78Pt .

* **All 3 known d9 elements violate Madelung rule**: 29Cu, 47Ag, 79Au.

The statistics for f-shell is only 2 elements, therefore we may give only some hints about their behaviour:

* f1, f2 and f8 always violate Madelung rule (true for both elements in these cells): 57La & 89Ac, 58Ce & 90Th and 64Gd & 96Cm.

* f6, f9, f10, f11, f12 and f13 never violate Madelung rule (true for both elements in these cells): 62Sm & 94Pu, 65Tb & 97Bk, 66Dy & 98 Cf, 67Ho & 99Es, 68Er & 100Em, 69Tm & 101Md.

Thou, as there are only 2-4 elements for each cell no strict rules can be defined on their basis. However, d2, d8, f6 and f8 elements are explained by 8-rule, discussed later.

Let's consider the order of the violations appearance.

Both during 1st appearance of d and f shell three elements violate Madelung rule. Both during 2nd appearance of d and f shell six elements violate Madelung rule. Due to this very interesting fact we may formulate the following suggestion.

During the first appearance of d-shell elements with 1, 2 and 6 electron deficiency violate Madelung rule. During the first appearance of f-shell elements with 1, 2 electrons and with 6 electron deficiency violate Madelung rule.

During the second appearance of d-shell elements with 1, 2, 3 and 4 and also 6 and 7 electron deficiency violate Madelung rule. By the same pattern during the second appearance of f-shell elements with 1, 2, 3 and 4 electron violate Madelung rule, and also element with 6 electron deficiency. A special element is f5 (93Np) that has 5 electrons and the corresponding 9 electron deficiency.

Moreover the double violation happens at d8 element (2 electron deficiency) and the next double violation happens at f2 element (i.e. the element with 2 electrons!).

During the third appearance of d-shell elements with 1 and 2 electron deficiency violate Madelung rule. And during the fourth appearance of d-shell 1st element is known to have complex violation.

Therefore we see antisymmetry of f and d shell, when violation status of d-elements with 1-4 electron deficiency is copied to the element same number of electrons in the f-shell of the same appearance queue; element 6 is copied without changes and element with 7 electron deficiency in d-shell (it has 3 electrons) relates somehow to f-element with 5 electrons and 9 electron deficiency.

Therefore we may have another clue on first appearance of g-shell elements violation. According to asymmetry, discussed in the paragraph 4 we may guess that g17 and g16 elements will violate Madelung rule and g12 (as element with 6 electron deficiency) is another candidate for Madelung rule violation.

## 3. Madelung rule violation statistics

As soon as we have some strict rules for Madelung rule violation given in previous chapter we may closely consider other elements statistics. We can make an assumption that violation events are not completely random, but follow some kind of a statistical dependency.

The first idea to investigate is usage of fig.3a for prognosis of other elements electron structure. Thou there are only 2-4 elements of this data for d and f elements per cell and it cannot be used to predict g-elements electron structure (g-elements start from 121th and are very close to contemporary experimental achievements).

Therefore anticipating some interconnection between electron number end electron deficiency we propose the following model.

Let us write down the d and f elements that follow Madelung rule as '+' and that violate Madelung rule as '-'. Sorting them by order of element appearance in a shell and the same in an reversed order, we obtain a Table 1-2. First two lines of the table give the total of corresponding f and d elements.

One note should be made here that we do not consider here elements that follow the extended set of rules given above (colored in yellow). We consider them as 'always true' in our model.

Table 1. Madelung rule violation sorted by electron number

| | 1 | 2 | 3 | 4 | 5 | 6 | 7 | 8 | 9 | 10 | 11 | 12 | 13 | 14 |
|---|---|---|---|---|---|---|---|---|---|---|---|---|---|---|
| **+** | **3** | **4** | **3** | **2** | **1** | **4** | **2** | **-** | **2** | **2** | **2** | **2** | **2** | **-** |
| **-** | **3** | **2** | **2** | **3** | **1** | **1** | **1** | **5** | **3** | **-** | **-** | **-** | **-** | **-** |
| **d(+)** | 3 | 4 | 2 | 1 | 3 | 2 | 2 | - | - | 3 | | | | |
| **d(-)** | 1** | - | 1 | 2 | - | 1 | 1 | 3* | 3 | - | | | | |
| **f(+)** | - | - | 1 | 1 | 1 | 2 | 2 | - | 2 | 2 | 2 | 2 | 2 | 2 |
| **f(-)** | 2 | 2* | 1 | 1 | 1 | - | - | 2 | - | - | - | - | - | - |

Table 2. Madelung rule violation sorted by electron deficiency.

| | -13 | -12 | -11 | -10 | -9 | -8 | -7 | -6 | -5 | -4 | -3 | -2 | -1 | 0 |
|---|---|---|---|---|---|---|---|---|---|---|---|---|---|---|
| **+** | **-** | **-** | **1** | **1** | **4** | **6** | **2** | **1** | **2** | **4** | **4** | **2** | **2** | **-** |
| **-** | **2** | **2** | **1** | **1** | **2** | **-** | **1** | **4** | **-** | **1** | **1** | **3** | **3** | **-** |
| **d(+)** | | | | | 3 | 4 | 2 | 1 | 3 | 2 | 2 | - | - | 3 |
| **d(-)** | | | | | 1** | - | 1 | 2 | - | 1 | 1 | 3* | 3 | - |
| **f(+)** | - | - | 1 | 1 | 1 | 2 | 2 | - | 2 | 2 | 2 | 2 | 2 | 2 |
| **f(-)** | 2 | 2* | 1 | 1 | 1 | - | - | 2 | - | - | - | - | - | - |

* one double violation considered
** one complex violation considered

The statistics, presented in Table 1 doesn't present some kind of a rule, but rather another presentation of fig.3. It is clearly seen that f(+) and f(-) lines are exactly the same in both tables, and d(+) and d(-) lines are shifted by 4 elements (it's quiet obvious).

We may note that all 5 elements with 8 electrons violate Madelung rule, and all 6 elements with 8 electron deficiency always obey Madelung rule.

Let us suggest that this concludes the final rule for our extended rules set:

**8-rule**: **every element with 8 electrons violates Madelung rule and every element with 8 electron deficiency obeys Madelung rule.**

Let's note that there is a controversy over 28Ni electron structure, which is d8 element, and according to 8-rule it should also violate Madelung rule (we consider Ni violating Madelung rule in this paper). Moreover, 8-rule, again, reveals some symmetry in Madelung rule violation statistics.

Probably, there is also some kind of a rule for 10 electrons and for 5 electrons deficiency which always obey Madelung rule, but out of 5 elements for each case 3 of them obey to abovegiven extended set of rules, therefore it's a question of concern.

However, such idea may be further developed and next step will be building of elements statistic model. The method proposed in this paper is the following.

We find the element's number of electrons in non-closed shells. Then we find its electron deficiency. Then we add the rows and columns and obtain separately the number of elements that follow and violate the Madelung rule. Dividing the number of elements that follow the Madelung rule by total number of elements processed we obtain the probability of obeying Madelung rule by a specific element. I.e. the formula will be the following:

$$p = (n_{1+} + n_{2+}) / (n_{1+} + n_{1-} + n_{2+} + n_{2-}) \quad (1)$$

where $n_{1+}$ and $n_{1-}$ are '+' and '-' values taken from table 1 and $n_{2+}$ and $n_{2-}$ values are '+' and '-' values taken from table 2 for the corresponding element number of electrons in an unclosed electron shell and electron deficiency respectively.

For example for d3 element the calculation scheme will be the following. As it has 3 electrons in the non-closed electron shell based on table 1 $n_{1+} = 3$ and $n_{1-} = 2$. As for its electron shell it electron deficiency is 7 electrons. Therefore we take from table 2 other pair of $n_{2+}$ and $n_{2-}$ values: 2 and 1 respectively.

And finally dividing all '+' values (5) by total '+' and '-' values (8) we obtain 0.625 ('experimental' value is 0.667) which is the probability of the element not violating the Madelung rule. And vice versa 1 - 0.625 = 0.375 is probability of Madelung rule violation by d3 element.

The results of such calculations are given in fig.4. As obvious, it includes extended rules set from previous paragraph and 8-rule from this paragraph.

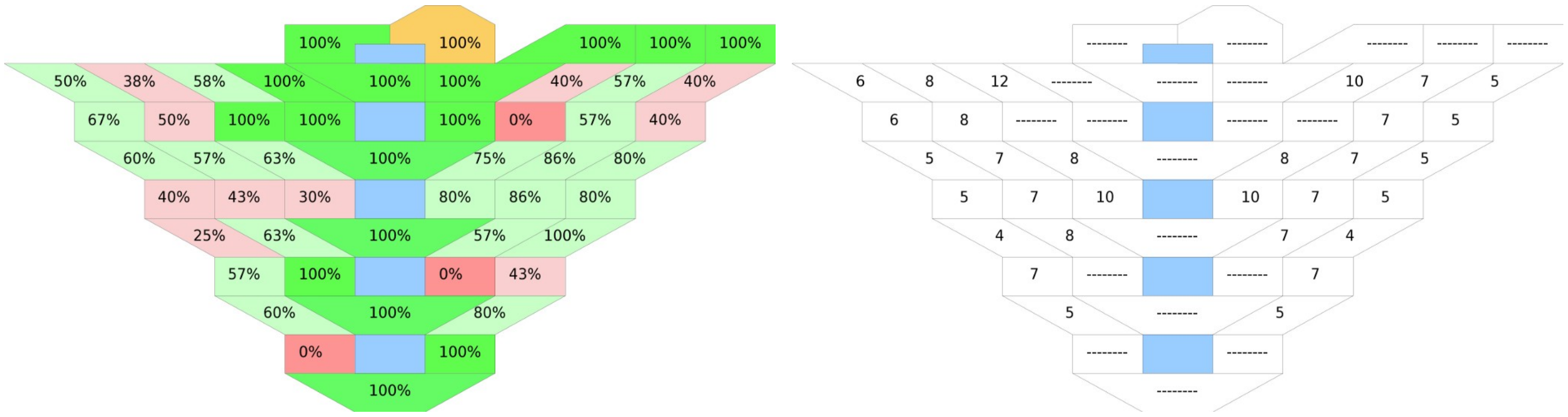

(a) (b)

Fig.4. Statistics for elements obeying the Madelung rule (a) and the number of elements included in the statistics (b) for every cell (elements determined by extended set of rules are not included); elements with probability of 50% and more are highlighted with green color and red color is used otherwise; intense colors denote extended set of rules.

Comparing fig.4 and fig.3 we can calculate the correlation coefficients for the model and experimental data 0.92 for d elements and 0.76 for f elements with total (mean) correlation 0.82.

Based on high correlation coefficient between model and experimental data both for d and f elements we can suggest that the correlation coefficient will be of the same order for yet unknown g-elements, however, subject to high statistical error as seen at fig.4(b).

Let's note that 121-124 (i.e. g1-g4) and 137-134 (i.e. g15-g17) elements are determined by no table 1 and table 2 statistics overlap, therefore the model is not absolutely correct here.

## 4. Anti-symmetrical correlation model

Looking both at fig.3(a) and fig.4(a) we may note some anti-symmetry in probability for d, f and g elements (especially well seen in colors). Therefore it is interesting to build an asymmetrical variant of the Madelung rule violation model that sum of the opposite cells will be equal to 100% by evenly increasing or reducing corresponding probabilities. The result may be found in fig.5.

In all cases except for d3 & d7 and f3 & f11 element pair the required change did not exceed 1/n, where n is the number of elements available in statistics (fig.4(b)).

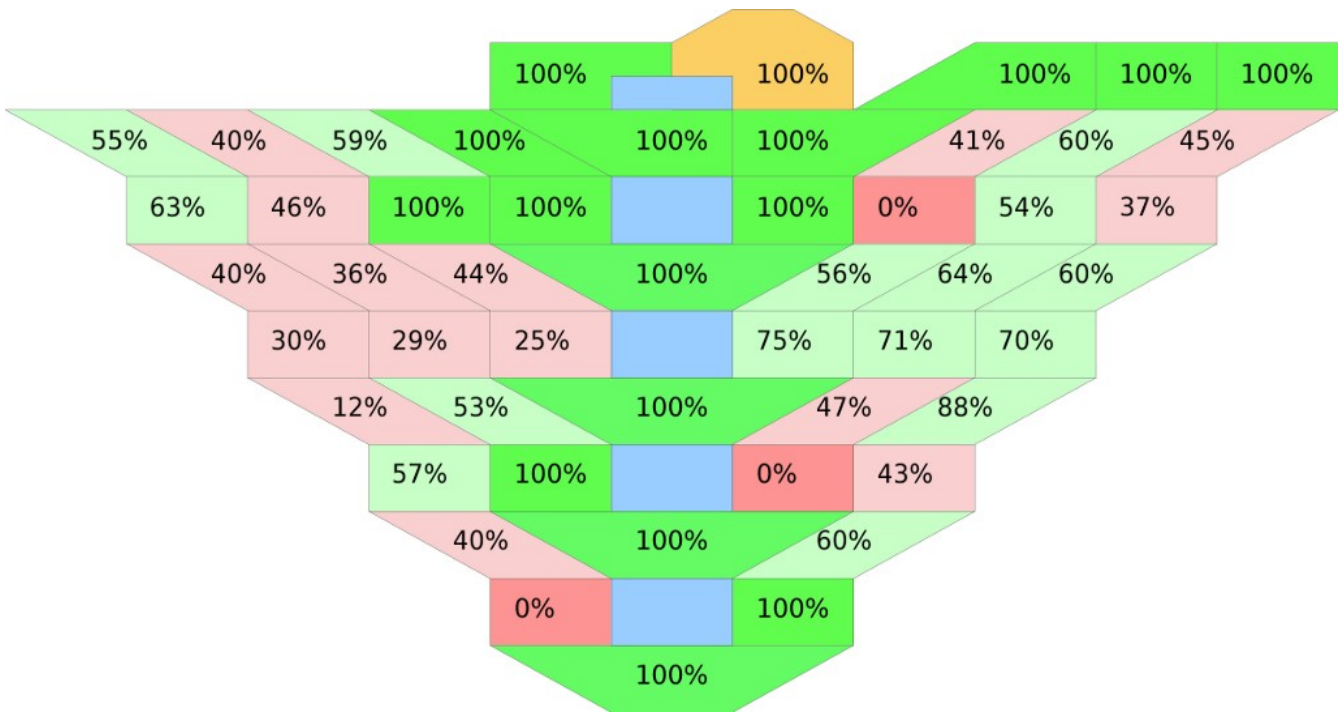

Fig.5. The anti-symmetrical correlation model (colors are equal to fig.4).

However the correlation coefficients for this model and experimental data reduces and is found to be 0.89 for d-elements, and 0.71 for f-elements and 0.77 mean correlation for both d and f elements. This may be the consequence of more complex symmetry-antisymmetry law discussed in paragraph 1.

Therefore later we consider non-corrected model (fig.4).

## 5. Prediction of superheavy elements electron structure

According to the obtained results (fig.4a) we can propose electron structure of yet uninvestigated superheavy elements both synthesized and yet unsynthesized.

Some of these elements will obey the Madelung rule according to extended rules. And others will have some probability of obeying the Madelung rule. Finally, some elements may be predicted to violate or obey the Madelung rule according to additional data.

The results for elements 105 to 138th are presented in table 3. There are 5 d-elements and 14 g-elements with two possible electron shell structure (outside extended set of rules). Let's note that we do not present here zero probabilities, however treat such results as probable due to probability determination error.

Among the known elements not predefined by the extended set of rules 8 of 19 (42.1%) d-elements and 6 of 20 (30%) f-elements have simple Madelung rule violation. Next, double violation is 1 of 19 (5.2%) d-elements and 1 of 20 (5%) f-elements (total 2 of 39, i.e. 5.1%). And finally 1 element of 39 (2.6%) has complex violation.

In Table 3 there are 19 non-strict determined elements, therefore there is around 63% probability that among them an element will have double violation, and consequently 38% that two elements will have double violation. The most probable candidates are 110th or 111th elements because both are preceded by simple violating elements.

There is also an additional 39% chance of complex violation in this 19 elements.

As seen in experimental data there are no complex or double violations at first shell appearance (see paragraph 1), therefore according to such considerations g-shell should be free of non-simple violations. The same considerations propose that there should be only three elements at g-shell that have simple Madelung rule violation.

Table 3. Predicted electron structure of superheavy elements: the more probable electron shell structure is given in bold and the less probable is grayed.

| Element | Type | Predicted electron shell structure | notes |
|---|---|---|---|
| 105 | d3 | **(63%) [Rn] $5f^{14}$ $6d^{3}$ $7s^{2}$**<br>(37%) [Rn] $5f^{14}$ $6d^{4}$ $7s^{1}$ | |
| 106 | d4 | (30%) [Rn] $5f^{14}$ $6d^{4}$ $7s^{2}$<br>**(70%) [Rn] $5f^{14}$ $6d^{5}$ $7s^{1}$** | |
| 107 | d5 | **[Rn] $5f^{14}$ $6d^{5}$ $7s^{2}$** | M-rule |
| 108 | d6 | **(80%) [Rn] $5f^{14}$ $6d^{6}$ $7s^{2}$**<br>(20%) [Rn] $5f^{14}$ $6d^{7}$ $7s^{1}$ | |
| 109 | d7 | **(75%) [Rn] $5f^{14}$ $6d^{7}$ $7s^{2}$**<br>(25%) [Rn] $5f^{14}$ $6d^{8}$ $7s^{1}$ | |
| 110 | d8 | **[Rn] $5f^{14}$ $6d^{9}$ $7s^{1}$** | 8-rule<br>double violation(?) |
| 111 | d9 | (40%) [Rn] $5f^{14}$ $6d^{9}$ $7s^{2}$<br>**(60%) [Rn] $5f^{14}$ $6d^{10}$ $7s^{1}$** | All other 3 known d9 elements violate Madelung rule<br>double violation(?) |
| 112 | d10 | **[Rn] $5f^{14}$ $6d^{10}$ $7s^{2}$** | C-rule |
| 113 | p1 | **[Rn] $5f^{14}$ $6d^{10}$ $7s^{2}$ $7p^{1}$** | SP-rule |
| 114 | p2 | **[Rn] $5f^{14}$ $6d^{10}$ $7s^{2}$ $7p^{2}$** | SP-rule<br>some studies suggest it is an inert gas |
| 115 | p3 | **[Rn] $5f^{14}$ $6d^{10}$ $7s^{2}$ $7p^{3}$** | SP-rule |
| 116 | p4 | **[Rn] $5f^{14}$ $6d^{10}$ $7s^2$ $7p^4$** | SP-rule |
| 117 | p5 | **[Rn] $5f^{14}$ $6d^{10}$ $7s^2$ $7p^5$** | SP-rule |
| 118 | p6 | **[Rn] $5f^{14}$ $6d^{10}$ $7s^2$ $7p^6$** | inert gas<br>SP-rule |
| 119 | s1 | **[Rn] $5f^{14}$ $6d^{10}$ $7s^2$ $7p^6$ $8s^1$** | SP-rule<br>this and later elements are yet unsynthesized |
| 120 | s2 | **[Rn] $5f^{14}$ $6d^{10}$ $7s^2$ $7p^6$ $8s^2$** | SP-rule<br>inert gas like He(?) (see paragraph 1) |
| 121 | g1 | (50%) [Rn] $5f^{14}$ $5g^1$ $6d^{10}$ $7s^2$ $7p^6$ $8s^2$<br>(50%) [Rn] $5f^{14}$ $6d^{10}$ $6f^1$ $7s^2$ $7p^6$ $8s^2$ * | ** |
| 122 | g2 | **(67%) [Rn] $5f^{14}$ $5g^2$ $6d^{10}$ $7s^2$ $7p^6$ $8s^2$**<br>(33%) [Rn] $5f^{14}$ $5g^1$ $6d^{10}$ $6f^1$ $7s^2$ $7p^6$ $8s^2$ * | ** |
| 123 | g3 | **(60%) [Rn] $5f^{14}$ $5g^3$ $6d^{10}$ $7s^2$ $7p^6$ $8s^2$**<br>(40%) [Rn] $5f^{14}$ $5g^2$ $6d^{10}$ $6f^1$ $7s^2$ $7p^6$ $8s^2$ * | ** |
| 124 | g4 | (40%) [Rn] $5f^{14}$ $5g^4$ $6d^{10}$ $7s^2$ $7p^6$ $8s^2$<br>**(60%) [Rn] $5f^{14}$ $5g^3$ $6d^{10}$ $6f^1$ $7s^2$ $7p^6$ $8s^2$ *** | ** |
| 125 | g5 | (25%) [Rn] $5f^{14}$ $5g^5$ $6d^{10}$ $7s^2$ $7p^6$ $8s^2$<br>**(75%) [Rn]$5f^{14}$ $5g^4$ $6d^{10}$ $6f^1$ $7s^2$ $7p^6$ $8s^2$ *** | |
| 126 | g6 | **(57%) [Rn] $5f^{14}$ $5g^6$ $6d^{10}$ $7s^2$ $7p^6$ $8s^2$**<br>(43%) [Rn] $5f^{14}$ $5g^5$ $6d^{10}$ $6f^1$ $7s^2$ $7p^6$ $8s^2$ * | |
| 127 | g7 | **(60%) [Rn] $5f^{14}$ $5g^7$ $6d^{10}$ $7s^2$ $7p^6$ $8s^2$**<br>(40%) [Rn] $5f^{14}$ $5g^6$ $6d^{10}$ $6f^1$ $7s^2$ $7p^6$ $8s^2$ * | |
| 128 | g8 | **[Rn] $5f^{14}$ $5g^9$ $6d^{10}$ $6f^1$ $7s^2$ $7p^6$ $8s^2$ *** | 8-rule |
| 129 | g9 | **[Rn] $5f^{14}$ $5g^9$ $6d^{10}$ $7s^2$ $7p^6$ $8s^2$** | M-rule |
| 130 | g10 | **[Rn] $5f^{14}$ $5g^{10}$ $6d^{10}$ $7s^2$ $7p^6$ $8s^2$** | 8-rule |
| 131 | g11 | **(80%) [Rn] $5f^{14}$ $5g^{11}$ $6d^{10}$ $7s^2$ $7p^6$ $8s^2$**<br>(20%) [Rn] $5f^{14}$ $5g^{10}$ $6d^{10}$ $6f^1$ $7s^2$ $7p^6$ $8s^2$ * | |
| 132 | g12 | (43%) [Rn] $5f^{14}$ $5g^{12}$ $6d^{10}$ $7s^2$ $7p^6$ $8s^2$<br>**(57%) [Rn] $5f^{14}$ $5g^{11}$ $6d^{10}$ $6f^1$ $7s^2$ $7p^6$ $8s^2$ *** | should violate Madelung rule according to symmetry considerations in paragraph 1 |
| 133 | g13 | **(100%) [Rn] $5f^{14}$ $5g^{13}$ $6d^{10}$ $7s^2$ $7p^6$ $8s^2$** | should obey Madelung rule according to 5 electron deficiency, see note in paragraph 3. |
| 134 | g14 | **(80%) [Rn] $5f^{14}$ $5g^{14}$ $6d^{10}$ $7s^2$ $7p^6$ $8s^2$**<br>(20%) [Rn] $5f^{14}$ $5g^{13}$ $6d^{10}$ $6f^1$ $7s^2$ $7p^6$ $8s^2$ * | |
| 135 | g15 | **(80%) [Rn] $5f^{14}$ $5g^{15}$ $6d^{10}$ $7s^2$ $7p^6$ $8s^2$**<br>(20%) [Rn] $5f^{14}$ $5g^{14}$ $6d^{10}$ $6f^1$ $7s^2$ $7p^6$ $8s^2$ * | ** |
| 136 | g16 | (40%) [Rn] $5f^{14}$ $5g^{16}$ $6d^{10}$ $7s^2$ $7p^6$ $8s^2$<br>**(60%) [Rn] $5f^{14}$ $5g^{15}$ $6d^{10}$ $6f^1$ $7s^2$ $7p^6$ $8s^2$** * | should violate Madelung rule according to symmetry considerations in paragraph 1 ** |
| 137 | g17 | (40%) [Rn] $5f^{14}$ $5g^{17}$ $6d^{10}$ $7s^2$ $7p^6$ $8s^2$<br>**(60%) [Rn] $5f^{14}$ $5g^{16}$ $6d^{10}$ $6f^1$ $7s^2$ $7p^6$ $8s^2$** * | should violate Madelung rule according to symmetry considerations in paragraph 1 ** |
| 138 | g18 | **[Rn] $5f^{14}$ $5g^{18}$ $6d^{10}$ $7s^2$ $7p^6$ $8s^2$** | C-rule<br>does not exist (?) |

* For g-elements the simple Madelung rule violation may be not correctly presented because behavior of g-elements electron shell is yet unknown, for details see paragraph 1. I.e. there is a possibility that the g-electron will move not to $6f^1$ but to $7d^1$ or, less probable, to $6h^1$. However in this case the probabilities of violation do not change, and the change will be in substitution of relative electron instead of $6f^1$.

** No table 1 and table 2 overlap. The statistics may be not correct.

Therefore extended set of rules enables us to exactly determine electron shell structure of 13 superheavy elements: 107, 112-120 and 128-130. And the 138th element may also be 14th here if it exists. For other elements we are able to show probability of Madelung rule violation (and its type).

## Conclusion

1. The main property of the proposed tetrahedron periodical table is convenient account for electron number and deficiency. In case there are 138 elements in the table we may consider it almost spatially-symmetrical.

2. A set of 4 extended rules was formulated for the Madelung rule:

**SP-rule**: s and p elements never violate Madelung rule.

**C-rule**: every closed-shell element obeys Madelung rule.

**M-rule**: half-filled electron shell elements never violate Madelung rule.

**8-rule**: every element with 8 electrons violates Madelung rule and every element with 8 electron deficiency obeys Madelung rule.

3. A statistical model was proposed to determine the probability of Madelung rule violation for the elements not determined by extended rules.

4. Probable electron structure of yet uninvestigated superheavy elements is proposed.